\documentclass[sigconf]{acmart}
\usepackage{balance}
\usepackage{multirow}
\def\BibTeX{{\rm B\kern-.05em{\sc i\kern-.025em b}\kern-.08emT\kern-.1667em\lower.7ex\hbox{E}\kern-.125emX}}
    
\copyrightyear{2019} 
\acmYear{2019} 
\setcopyright{acmcopyright}
\acmConference[SIGIR '19]{Proceedings of the 42nd International ACM SIGIR Conference on Research and Development in Information Retrieval}{July 21--25, 2019}{Paris, France}
\acmBooktitle{Proceedings of the 42nd International ACM SIGIR Conference on Research and Development in Information Retrieval (SIGIR '19), July 21--25, 2019, Paris, France}
\acmPrice{15.00}
\acmDOI{10.1145/3331184.3331235}
\acmISBN{978-1-4503-6172-9/19/07}

\begin{document}

\settopmatter{printacmref=true}
\fancyhead{}

% The "title" command has an optional parameter, allowing the author to define a "short title" to be used in page headers.
\title{Cross-Modal Interaction Networks for Query-Based Moment Retrieval in Videos}

% The "author" command and its associated commands are used to define the authors and their affiliations.
% Of note is the shared affiliation of the first two authors, and the "authornote" and "authornotemark" commands
% used to denote shared contribution to the research.
% \author{Zhu Zhang}
% \email{zhangzhu@zju.edu.cn}
% \affiliation{%

\author{Zhu Zhang}
\affiliation{%
  \institution{Zhejiang University}
  \streetaddress{Road 38 West Lake District}
  \city{Hangzhou}
  \country{China}}
\email{zhangzhu@zju.edu.cn}

\author{Zhijie Lin}
\affiliation{%
  \institution{Zhejiang University}
  \streetaddress{Road 38 West Lake District}
  \city{Hangzhou}
  \country{China}}
\email{linzhijie@zju.edu.cn}

\author{Zhou Zhao}
\authornote{Zhou Zhao is the corresponding author.}
\affiliation{%
  \institution{Zhejiang University}
  \streetaddress{Road 38 West Lake District}
  \city{Hangzhou}
  \country{China}}
\email{zhaozhou@zju.edu.cn}

\author{Zhenxin Xiao}
\affiliation{%
  \institution{Zhejiang University}
  \streetaddress{Road 38 West Lake District}
  \city{Hangzhou}
  \country{China}}
\email{alanshawzju@gmail.com}

% \affiliation{%

% \affiliation{%
 
% \affiliation{%

% \affiliation{%



%
% By default, the full list of authors will be used in the page headers. Often, this list is too long, and will overlap
% other information printed in the page headers. This command allows the author to define a more concise list
% of authors' names for this purpose.
\renewcommand{\shortauthors}{Zhang and Lin, et al.}

%
% The abstract is a short summary of the work to be presented in the article.
\begin{abstract}
Query-based moment retrieval aims to localize the most relevant moment in an untrimmed video according to the given natural language query. Existing works often only focus on one aspect of this emerging task, such as the query representation learning, video context modeling or multi-modal fusion, thus fail to develop a comprehensive system for further performance improvement. In this paper, we introduce a novel Cross-Modal Interaction Network (CMIN) to consider multiple crucial factors for this challenging task, including (1) the syntactic structure of natural language queries; (2) long-range semantic dependencies in video context and (3) the sufficient cross-modal interaction. Specifically, we devise a syntactic GCN to leverage the syntactic structure of queries for fine-grained representation learning,  propose a multi-head self-attention to capture long-range semantic dependencies from video context, and next employ a multi-stage cross-modal interaction to explore the potential relations of video and query contents. 
The extensive experiments demonstrate the effectiveness of our proposed method. Our core code has been released at https://github.com/ikuinen/CMIN.
\end{abstract}


% \ccsdesc[500]{Information systems~Video search}
%
% Keywords. The author(s) should pick words that accurately describe the work being
% presented. Separate the keywords with commas.

%
% A "teaser" image appears between the author and affiliation information and the body 
% of the document, and typically spans the page. 
%   \Description{Enjoying the baseball game from the third-base seats. Ichiro Suzuki preparing to bat.}

%
% This command processes the author and affiliation and title information and builds
% the first part of the formatted document.
\maketitle

\begin{figure}[t]
\centering
\includegraphics[width=0.5\textwidth]{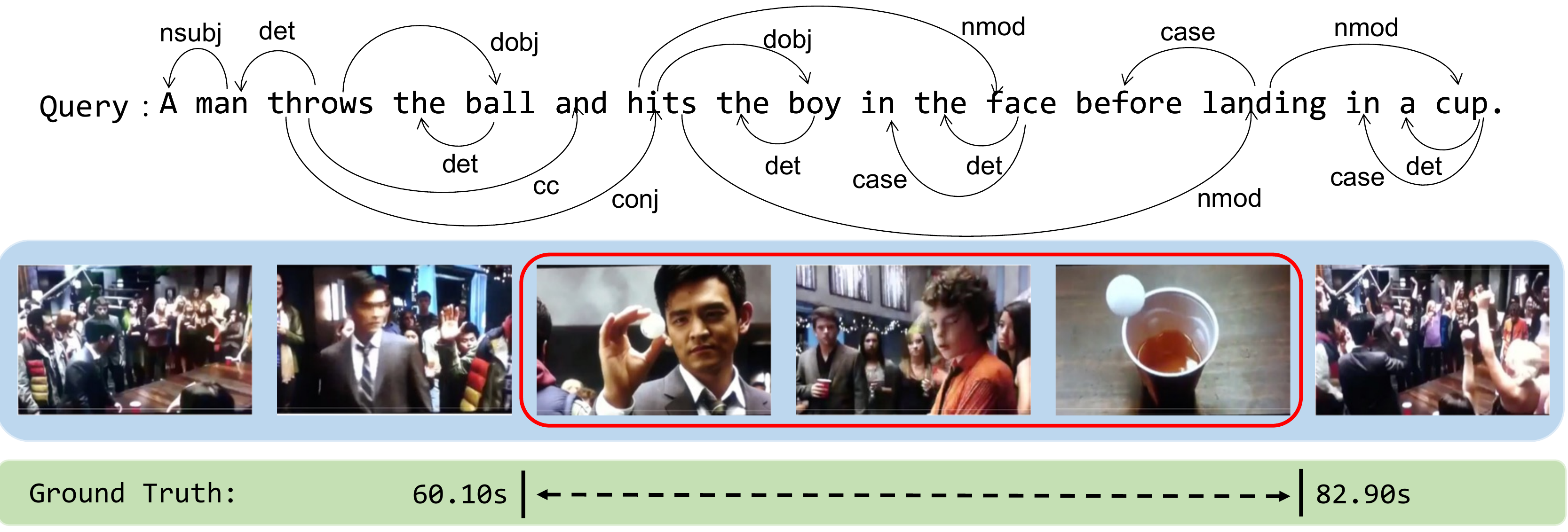}
\caption{Query-Based Moment Retrieval in Video}\label{fig:demo}
\end{figure}

\section{Introduction}
Multimedia information retrieval is an important topic in information retrieval systems.
Recently, query-based video retrieval~\cite{lin2014visual,xu2015jointly,otani2016learning} has been well-studied, which searches the most relevant video from large collections according to a given natural language query.
However, in practical applications, the untrimmed videos often contain multiple complex events that evolve over time, where a large part of video contents are irrelevant to the query and only a small clip satisfies the query description.
Thus, as a natural extension, the query-based moment retrieval aims to automatically localize the start and end boundaries of the target moment semantically corresponding to the given query within an untrimmed video.
Different from retrieving an entire video, moment retrieval offers more fine-grained temporal localization in a long video, which avoids manually searching for the moment of interests.

However, localizing the precise moment in a continuous, complicated video is more challenging than simply selecting a video from pre-defined candidate sets. As shown in Figure~\ref{fig:demo}, the query ~\lq\lq A man throws the ball and hits the boy in the face before landing in a cup\rq\rq\text{ } describes two successive actions, corresponding to complex object interactions within the video. 
Hence, the accurate retrieval of the target moment requires sufficient understanding of both video and query contents by cross-modal interactions.

Most existing moment retrieval works~\cite{gao2017tall,hendricks2018localizing,hendricks2017localizing,liu2018attentive,liu2018cross,chen2018temprally,xu2019multilevel} only focus on one aspect of this emerging task, such as the query representation learning~\cite{liu2018cross}, video context modeling~\cite{gao2017tall,liu2018attentive} and cross-modal fusion~\cite{chen2018temprally,xu2019multilevel}, thus fail to develop a comprehensive system to further improve the performance of query-based moment retrieval. In this paper, we consider multiple crucial factors for high-quality moment retrieval.

Firstly, the query description often contains causal temporal actions, thus it is fundamental and crucial to learn fine-grained query representations. Existing works generally adopt widely-used recurrent neural networks, such as GRU networks, to model natural language queries.
However, these approaches ignore the syntactic structure of queries. As shown in Figure~\ref{fig:demo}, the syntactic relation implies dependencies of word pairs, helpful for query semantic understanding.
Recently, the graph convolution networks (GCN) have been proposed to model the graph structure~\cite{kipf2017semi,velickovic2018graph}, including the visual relationship graph~\cite{yao2018exploring} and syntactic dependency graph~\cite{marcheggiani2017encoding}. Inspired by these works, we develop a syntactic GCN to exploit the syntactic structure of queries. Concretely, we first build the syntactic dependency graph as shown in Figure~\ref{fig:demo}, and then pass information along the dependency edges to learn syntactic-aware query representations. In detail, we consider the direction and label of dependency edges to adequately incorporate syntactic clues.

Secondly, the target moments of complex queries generally contain object interactions over a long time interval, thus there exist long-range semantic dependencies in video context. That is, each frame is not only relevant to adjacent frames, but also associated with distant ones.
Existing approaches often apply RNN-based temporal modeling~\cite{chen2018temprally}, or propose R-C3D networks to learn spatio-temporal representations from raw video streams~\cite{xu2019multilevel}. Although these methods are able to absorb contextual information for each frame, they still fail to build direct interactions between distant frames.
To eliminate the local restrictions, we propose a multi-head self-attention mechanism~\cite{vaswani2017attention} to capture long-range semantic dependencies from video context. The self-attention method can develop the frame-to-frame interaction at arbitrary positions and the multi-head setting ensures the sufficient understanding of complicated dependencies.

Thirdly, query-based moment retrieval requires the comprehensive reasoning of video and query contents, thus the cross-modal interaction is necessary for high-quality retrieval. 
Early approaches~\cite{gao2017tall,hendricks2017localizing,hendricks2018localizing} ignore this factor and only simply combine the query and moment features for correlation estimations.
Although recent methods~\cite{liu2018attentive,liu2018cross,chen2018temprally,xu2019multilevel} have developed a cross-modal interaction by widely-used attention mechanism, they still remain in the rough one-stage interaction, for example, highlighting the crucial context information of moments by the guidance of queries~\cite{liu2018attentive}.
Different from previous works, we adopt a multi-stage cross-modal interaction method to further exploit the potential relation of video and query contents. Specifically, we first adopt a normal attention method to aggregate syntactic-aware query representations for each frame, then apply a cross gate~\cite{feng2018video} to emphasize crucial contents and weaken inessential parts, and next develop the low-rank bilinear fusion to learn a cross-modal semantic representation.

In summary, the key contributions of this work are four-fold:
\begin{itemize}
\item We design a novel cross-modal interaction networks for query-based moment retrieval, which is a comprehensive system to consider multiple crucial factors of this challenging task: (1) the syntactic structure of natural language queries; (2) long-range semantic dependencies in video context and (3) the sufficient cross-modal interaction. 
\item We propose the syntactic GCN to leverage the syntactic structure of queries for fine-grained representation learning, and adopt a multi-head self-attention method to capture long-range semantic dependencies from video context. 
\item We employ a multi-stage cross-modal interaction to further exploit the potential relation of video and query contents, where an attentive aggregation method extracts relevant syntactic-aware query representations for each frame, a cross gate emphasizes crucial contents and a low-rank bilinear fusion method learn cross-modal semantic representations.
\item The proposed CMIN method achieves the state-of-the-art performance on ActivityCaptions and TACoS datasets.
\end{itemize}

The rest of this paper is organized as follows. We briefly review some related works in Section 2. In Section 3, we introduce our proposed method. We then present a variety of experimental results in Section 4. Finally, Section 5 concludes this paper.

\section{Related Work}
In this section, we briefly review some related works on image/video retrieval, temporal action localization and query-based moment retrieval.
\subsection{Image/Video Retrieval}
Given a set of candidate images/videos and a natural language query, image/video retrieval aims to select the image/video that matches this query.
Karpathy et al.~\cite{karpathy2015deep} propose a deep visual-semantic alignment (DVSA) model for image retrieval, which uses the BiLSTM to encode query features and R-CNN detector~\cite{girshick2014rich} to extract object representations.
Sun et al.~\cite{sun2015automatic} advise an automatic visual concept discovery algorithm to boost the performance of image retrieval.
Moreover, Hu et al.~\cite{hu2016natural} and Mao et al~\cite{mao2016generation} regard this problem as natural language object retrieval.
As for video retrieval, some methods~\cite{otani2016learning,xu2015jointly} incorporate deep video-language embeddings to boost retrieval performance, similar to the image-language embedding approach~\cite{socher2014grounded}.
And Lin et al.~\cite{lin2014visual} first parse the query descriptions into a semantic graph and then match them to visual concepts in videos.
Different from these works, query-based moment retrieval aims to localize a moment within an untrimmed video, which is more challenging than simply selecting a video from pre-defined candidate sets.

\begin{figure*}[t]
\centering
\includegraphics[width=1\textwidth]{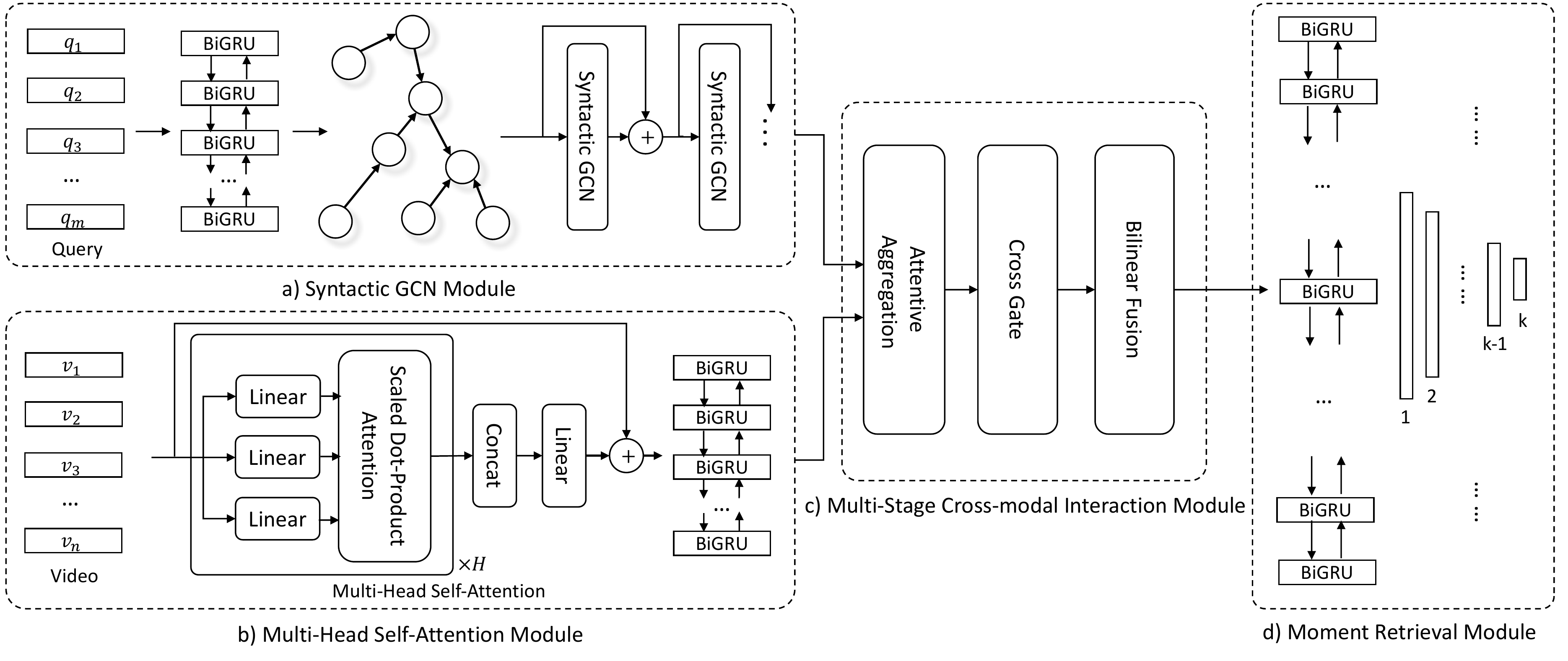}
\caption{The Framework of Cross-Modal Interaction Networks for Query-Based Moment Retrieval. (a) The syntactic GCN module leverages the syntactic structure to learn syntactic-aware query representations. (b) The multi-head self-attention module captures long-range semantic dependencies from context. (c) The multi-stage cross-modal interaction module explores the intrinsic relations between video and query. (d) The moment retrieval module localizes the boundaries of target moments.}\label{fig:framework}
\end{figure*}

\subsection{Temporal Action Localization}
Temporal action localization is a challenging task to localize action instances in an untrimmed video.
Shou et al.~\cite{shou2016temporal} develop three segment-based 3D ConvNets with localization loss to explicitly explore the temporal overlap in videos. 
Singh et al.~\cite{singh2016multi} propose a multi-stream bi-directional RNN with two additional streams on motion and appearance to achieve the fine-grained action detection.

To leverage the context structure of actions, Zhao et al.~\cite{zhao2017temporal} advise a structured segment network to model the structure of action instances by a structured temporal pyramid.
And Chao et al.~\cite{chao2018rethinking} boost the action localization performance by imitating the Faster RCNN object detection framework~\cite{ren2015faster}. 
Although these works have achieved promising performance, they still are limited to a pre-defined list of actions.
And query-based moment retrieval tackles this problem by introducing the natural language query.

\subsection{Query-Based Moment Retrieval}
Query-based moment retrieval is to detect the target moment depicting the given natural language query in an untrimmed video.
Early works study this task in constrained settings, including the fixed spatial prepositions~\cite{tellex2009towards,lin2014visual}, instruction videos~\cite{alayrac2016unsupervised,sener2015unsupervised,song2016unsupervised} and ordering constraint~\cite{bojanowski2015weakly,tapaswi2015book2movie}. 
Recently, unconstrained query-based moment retrieval has attracted a lot of attention~\cite{hendricks2017localizing, gao2017tall,liu2018attentive,hendricks2018localizing,liu2018cross,xu2019multilevel,chen2018temprally}.
These methods are mainly based on a sliding window framework, which first samples candidate moments and then ranks these moments.
Hendricks et al.~\cite{hendricks2017localizing} propose a moment context network to integrate global and local video features for natural language retrieval, and the subsequent work~\cite{hendricks2018localizing} considers the temporal language by explicitly modeling the context structure of videos.
Gao et al.~\cite{gao2017tall} develop a cross-modal temporal regression localizer to estimate the alignment scores of candidate moments and textual query, and then adjust the boundaries of high-score moments.
With the development of attention mechanism in the field of vision and language interaction~\cite{Anderson2018BottomUpAT,zhao2018open}, Liu et al.~\cite{liu2018attentive} advise a memory attention to emphasize the visual features and simultaneously utilize the context information. And similar attention strategy~\cite{liu2018cross} is designed to highlight the crucial part of query contents.
From the holistic view, Chen et al.~\cite{chen2018temprally} capture the evolving fine-grained frame-by-word interactions between video and query.
Xu et al.~\cite{xu2019multilevel} introduce a multi-level model to integrate visual and textual features earlier and further re-generate queries as an auxiliary task.

Unlike these previous methods, we propose a novel cross-modal interaction network to consider three critical factors for query-based moment retrieval, including the syntactic structure of natural language queries, long-range semantic dependencies in video context and the sufficient cross-modal interaction.

\section{Cross-Modal Interaction Networks}
As Figure~\ref{fig:framework} illustrates, our cross-modal interaction networks consist of four components: 1) the syntactic GCN module leverages the syntactic structure to enhance the query representation learning; 2) the multi-head self-attention module captures long-range semantic dependencies from video context;
3) the multi-stage cross-modal interaction module aggregates syntactic-aware query representations for each frame, emphasizes crucial contents and learns cross-modal semantic representations; 4) the moment retrieval module finally localizes the boundaries of target moments.

\subsection{Problem Formulation}
We present a video as a sequence of frames ${\bf v} = \{{\bf v}_{i}\}_{i=1}^{n} \in V$, where ${\bf v}_i$ is the feature of the $i$-th frame and $n$ is the frame number of the video. 
Each video is associated with a natural language query, denoted by ${\bf q} = \{{\bf q}_{i}\}_{i=1}^{m} \in Q$, where ${\bf q}_i$ is the feature of the $i$-th word and $m$ is the word number of the query. The query description corresponds to a target moment in the untrimmed video and we denote the start and end boundaries of the target moment by ${\bf \tau} = (s, e) \in A$.
Thus, given the training set $\{V, Q, A\}$, our goal is to learn the cross-modal interaction networks to predict the boundary ${\hat \tau} = ({\hat s}, {\hat e})$ of the most relevant moment during inference.

\subsection{Syntactic GCN Module}
In this section, we introduce the syntactic GCN module based on a syntactic dependency graph. By passing information along the dependency edges between relevant words, we learn syntactic-aware query representations for subsequent cross-modal interactions. 

We first extract word features for the query using a pre-trained Glove word2vec embedding~\cite{pennington2014glove}, denoted by ${\bf q} = ({\bf q }_1, {\bf q }_2,\ldots, {\bf q }_m)$, where ${\bf q}_i$ is the feature of the $i$-th word. After that, we develop a bi-directional GRU networks (BiGRU) to learn the query semantic representations. The BiGRU networks incorporate contextual information for each word by combining the forward and backward GRU~\cite{chung2014empirical} . Sepcifically, we input the sequence of word features to the BiGRU networks, and obtain the contextual representation of each word, given by
\begin{equation}
\begin{split}
&{\bf h}^{f}_{i} = {\rm GRU}^f_q({\bf q}_{i},{\bf h}^{f}_{i-1}),   \\
&{\bf h}^{b}_{i} = {\rm GRU}^b_q({\bf q}_{i},{\bf h}^{b}_{i+1}),   \\
&{\bf h}^{q}_{i} = [{\bf h}^{f}_{i};{\bf h}^{b}_{i}],
\end{split}
\end{equation}
where ${\rm GRU}^f_q$ and ${\rm GRU}^b_q$ represent the forward and backward GRU networks, respectively. And the contextual representation ${\bf h}^{q}_{i}$ is the concatenation of the forward and backward hidden state at the $i$-th step. Thus, we get the query semantic representations ${\bf h}^q = ({\bf h }^q_1, {\bf h }^q_2,\ldots, {\bf h }^q_m)$.

Although the BiGRU networks have encoded temporal context of word sequences, they still ignore the syntactic information of natural language, which implies underlying dependencies between word pairs. So we then advise the syntactic graph convolution networks to leverages the syntactic dependencies for better query understanding.
We first build the syntactic dependency graph by an NLP toolkit, where each word is regarded as a node and each dependency relation is presented as a directed edge. 
Formally, we denote a query by a graph $\mathcal{G = (V,E)}$, where the node set $\mathcal{V}$ contains all words and edge set $\mathcal{E}$ contains all directed syntactic dependencies of word pairs. Note that we add the self-loop for each node into the edge set. 
As the dependency relations have different types, the directed edges also correspond to different labels, where the self-loop is given a unique label.
The original GCN regard the syntactic dependency graph as an undirected graph, denoted by
\begin{eqnarray}
{\bf g}_i^1 = {\rm ReLU}\left(\sum_{j \in {\mathcal N}(i)}{\bf W}^g {\bf h}_j^q+{\bf b}^g\right) 
\end{eqnarray}
where ${\bf W}^g$ is the transformation matrix, ${\bf b}^g$ is the bisa vector and ${\rm ReLU}$ is the rectified linear unit. The ${\mathcal N}(i)$ represents the set of nodes with a dependency edge to node $i$ or from node $i$ (including self-loop). And the ${\bf h}_j^q$ is the original representation of node $j$ from the preceding modeling. 

\begin{figure}[t]
\centering
\includegraphics[width=0.45\textwidth]{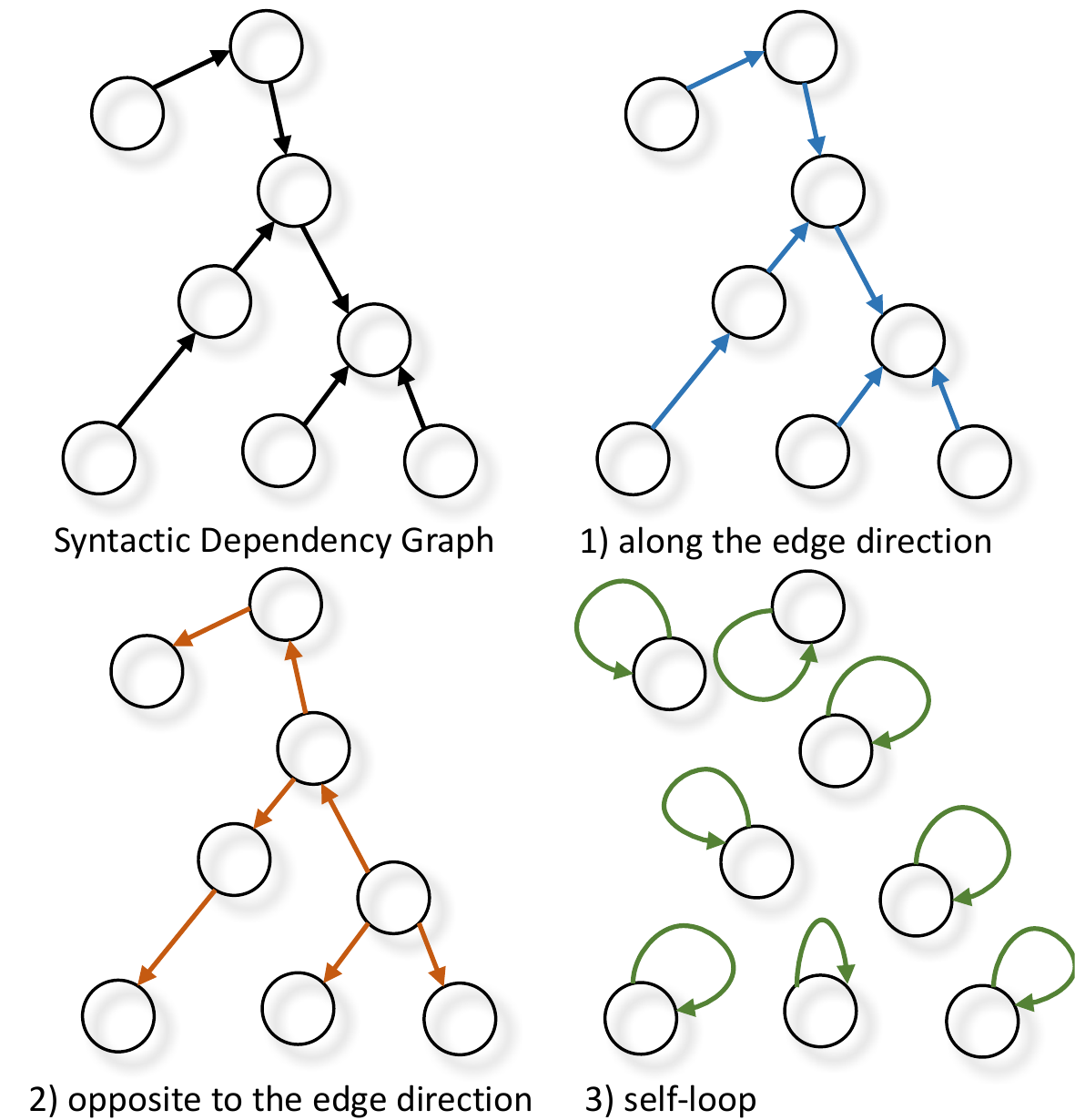}
\caption{Three Directions of Information Passing.}\label{fig:framework2}
\end{figure}

Although the original GCN enhances the word semantic representation by aggregating the clues from its neighbors, it fails to leverage the direction and label information of edges. Thus, we consider a syntactic GCN to exploit the directional and labeled dependency edges between nodes, given by
\begin{eqnarray}
{\bf g}_i^1 = {\rm ReLU}\left(\sum_{j \in {\mathcal N}(i)}{\bf W}^g_{dir(i,j)} {\bf h}_j^q+{\bf b}^g_{lab(i,j)}\right) 
\end{eqnarray}
where ${dir(i,j)}$ indicates the direction of edge $(i,j)$: 1) a dependency edge from node $i$ to $j$; 2) a dependency edge from node $j$ to $i$; 3) a self-loop if $i = j$.  
Since there is no reason to assume the information transmits only along the syntactic dependency arcs~\cite{marcheggiani2017encoding},  we also allow the information to transmit in the opposite direction of directed dependencies here. The Figure~\ref{fig:framework2} describes the three types of information passing, which correspond to three transformation matrix ${\bf W}^g_1$,${\bf W}^g_2$ and ${\bf W}^g_3$, respectively.
On the other hand, the $lab(i,j)$ represents the label of edge $(i,j)$ to select a distinct bias vector for each type of dependencies.
Next, we employ a residual connection~\cite{he2016deep} to keep the original representation of each node, given by
\begin{eqnarray}
{\bf o}_i^1 = {\bf g}_i^1 + {\bf h}_i^q.
\end{eqnarray}
Furthermore, we stack a multi-layer syntactic GCN to adequately explore the syntactic structure as follows.
\begin{equation}
\left\{
             \begin{array}{lr}
             {\bf g}^1 = {\rm synGCN}({\bf h}^q), \ {\bf o}^1 = {\bf g}^1 + {\bf h}^q \\
             {\bf g}^2 = {\rm synGCN}({\bf o}^1), \ {\bf o}^2 = {\bf g}^2 + {\bf o}^1 \\
             ... \\
             {\bf g}^l = {\rm synGCN}({\bf o}^{l-1}), \ {\bf o}^l = {\bf g}^l + {\bf o}^{l-1} 
             \end{array}
\right.
\end{equation}
By the syntactic GCN with $l$ layers, we obtain the syntactic-aware query representation ${\bf o}^l = ({\bf o}^l_1, {\bf o}^l_2, \ldots, {\bf o}^l_m)$.

\subsection{Multi-Head Self-Attention Module}
In this section, we present the multi-head self-attention module to capture long-range semantic dependencies from video context. By the self-attention method, each frame is able to interact not only with adjacent frames but also with distant ones. And the multi-head setting is beneficial to sufficiently understand the complicated dependencies.

We first extract frame features from the untrimmed video by a pre-trained 3D-ConvNet~\cite{tran2015learning}, denoted by ${\bf v} = ({\bf v}_1, {\bf v}_2,\ldots, {\bf v}_n)$. The ${\bf v}_i$ is the visual feature of the $i$-th frame.
We then introduce the multi-head self-attention based on the scaled dot-product attention, which originally is proposed in the field of machine translation~\cite{vaswani2017attention}.

\textbf{Scaled dot-product attention.}  We assume the input of the scale dot-product attention is a sequence of queries $\overline{\bf Q} \in \mathbb{R}^{d_k \times n_q}$, keys $\overline{\bf K} \in \mathbb{R}^{d_k \times n_k} $ and values $\overline{\bf V} \in \mathbb{R}^{d_v \times n_k}$, where $n_q$ , $n_k$ and $n_k$ represent the number of queries, keys and values, and $d_k$, $d_k$  and $d_v$ are respective dimensions. 
The scaled dot-product attention is then calculated by
\begin{eqnarray}
{\rm Attention}(\overline{\bf Q},\overline{\bf K},\overline{\bf V}) = {\rm Softmax}(\frac{\overline{\bf Q}^{\top}\overline{\bf K}}{\sqrt{d_k}})\overline{\bf V}^{\top},
\end{eqnarray}
where the $ {\rm Softmax}$ operation is performed on every row. The values are aggregated for each query according to the dot-product score between the query and the corresponding key of values.

\textbf{Multi-head attention.} The multi-head attention consists of $H$ paralleled scaled dot-product attention layers. For each independent attention layer, the input queries, keys and values are linearly projected to $d_k$, $d_k$ and $d_v$ dimenstions. Concretely, the result of multi-head attention is given by
\begin{equation}
\begin{split}
&{\rm MultiHead}(\overline{\bf Q},\overline{\bf K},\overline{\bf V}) = {\bf W}^O{\rm Concat(head_1,\ldots, head_H)}  \\
&{\rm where \ head_i = Attention({\bf W}_i^Q\overline{\bf Q}, {\bf W}_i^K\overline{\bf K}, {\bf W}_i^V\overline{\bf V})} 
\end{split}
\end{equation}
where ${\bf W}_i^Q \in \mathbb{R}^{d_k \times d_1}$, ${\bf W}_i^K \in \mathbb{R}^{d_k \times d_2}$, ${\bf W}_i^V \in \mathbb{R}^{d_v \times d_3}$, ${\bf W}^O \in \mathbb{R}^{d_4 \times Hd_v}$ are linear projection matrices. The $d_1$, $d_2$ ,$d_3$ are the initial input dimensions and $d_4$ is the output dimension. 

When the queries $\overline{\bf Q}$, keys $\overline{\bf K}$ and values $\overline{\bf V}$ are set to the same video feature matrix ${\bf V} = [{\bf v}_1; {\bf v}_2;\ldots; {\bf v}_n] \in \mathbb{R}^{d \times n} $ and the input dimension is equal to the output dimension, we get a multi-head self-attention method.
Based on it, we obtain the self-attentive video representation ${\bf V}^s = [{\bf v}^s_1; {\bf v}^s_2;\ldots; {\bf v}^s_n]$, given by
\begin{eqnarray}
{\bf V}^s = {\rm MultiHead}({\bf V},{\bf V},{\bf V}) + {\bf V},
\end{eqnarray}
where a residual connection is applies similar to syntactic GCN. Here we establish frame-to-frame correlation among video sequences and the multi-head setting allows the attention operation to aggregate information from different representation subspaces.
But the temporal modeling is still critical for video semantic understanding, we cannot only depend on the multi-head self-attention and ignore contextual representation learning.
Hence, we next employ another BiGRU to learn the self-attentive video semantic representations ${\bf h}^v = ({\bf h }^v_1, {\bf h }^v_2,\ldots, {\bf h }^v_n)$.

\subsection{Multi-Stage Cross-Modal Interaction Module}
In this section, we introduce the cross-modal interaction module to exploit the potential relations of video and query contents, which consists of the attentive aggregation, cross-gated interaction and low-rank bilinear fusion. 

\textbf{Attentive aggregation.} Given the syntactic-aware query representations ${\bf o}^l = ({\bf o}^l_1, {\bf o}^l_2, \ldots, {\bf o}^l_m)$ and self-attentive video semantic representations ${\bf h}^v = ({\bf h }^v_1, {\bf h }^v_2,\ldots, {\bf h }^v_n)$, we apply a typical attention mechanism to aggregate the query clues for each frame. Concretely, we first compute the attention score between each pair of frame and word, and obtain a video-to-query attention matrix $M \in \mathbb{R}^{n \times m}$. The attention score of the $i$-th frame and $j$-th word is given by
\begin{eqnarray}
M({\bf h}_{i}^{v}, {\bf o}_{j}^{l})= {\bf w}^{\top}tanh({\bf W}^{m}_{1} {\bf h}_{i}^{v}+ {\bf W}^{m}_{2}{\bf o}_{j}^{l}+{\bf b}^{m}), 
\end{eqnarray}
where $ {\bf W}_{1}^{m}$, $ {\bf W}_{2}^{m}$ are parameter matrices, $ {\bf b}^{m}$ is the bias vector and the ${\bf w}^{\top}$ is the row vector. We then apply the softmax operation for each row of $M$, given by
\begin{eqnarray}
M^{row}_{ij} = \frac{exp(M_{ij})}{\sum_{k=1}^m exp(M_{ik})},
\end{eqnarray}
where $M^{row}_{ij}$ represents the correlation of the $i$-th frame and $j$-th word.
Next, we extract the crucial query clues for each frame based on $M^{row}$, given by
\begin{eqnarray}
& {\bf h}^{s}_{i} = \sum_{j=1}^{m} M^{row}_{ij} {\bf o}_{j}^{l},   
\end{eqnarray}
where the ${\bf h}^{s}_{i}$ represents the aggregated query representation relevant to the $i$-th frame. 

\textbf{Cross-gated interaction.} With the aggregated query representation ${\bf h}^{s}_{i}$  and frame semantic representation ${\bf h}^{v}_{i}$, we then apply a cross gate~\cite{feng2018video} to emphasize crucial contents and weaken inessential parts. 
In the cross gate method, the gate of query representation depends on the frame representation, and meanwhile the frame representation is also gated by its corresponding query representation, denoted by
\begin{equation}
\begin{split}
&{\bf g}_i^v = \sigma({\bf W}^v{\bf h}^v_i + {\bf b}^v), \\
&{\bf \widetilde h}^s_i = {\bf h}^s_i \odot {\bf g}_i^v, \\
&{\bf g}_i^s = \sigma({\bf W}^s{\bf h}^s_i + {\bf b}^s), \\
&{\bf \widetilde h}^v_i = {\bf h}^v_i \odot {\bf g}_i^s,
\end{split}
\end{equation}
where $ {\bf W}^{v}$, $ {\bf W}^{s}$ are parameter matrices, ${\bf b}^{v}$ and ${\bf b}^{s}$ are the bias vectors, $\sigma$ is the sigmoid function, and $\odot$ represents element-wise multiplication.
If the aggregated query representation ${\bf h}^s_i$ is irrelevant to the frame semantic representation ${\bf h}^{v}_{i}$, both the two representations are filtered to decrease their influences on subsequent networks. On the contrary, the cross gate can further enhance the effects of relevant frame-query pairs.

\textbf{Bilinear fusion.} After the attentive aggregation and cross gate, we propose a low-rank bilinear fusion method~\cite{kim2017hadamard} to further exploit the cross-modal interaction between ${\bf \widetilde h}^s_i$ and ${\bf \widetilde h}^v_i$.
The original bilinear fusion method is written by
\begin{eqnarray}
f_{ij} = {\bf \widetilde h}^{v^{\top}}_i {\bf W}^f_j {\bf \widetilde h}^s_i + {\bf b}^f_j,
\end{eqnarray}
where $f_{ij}$ represents the $j$-th dimension of the bilinear output at the time step $i$, and the ${\bf f}_i$ is the fusion result of ${\bf \widetilde h}^s_i$ and ${\bf \widetilde h}^v_i$. the $ {\bf W}^{f}_{j}$, $ {\bf b}^{f}_{j}$ are the parameter matrix and the bias vectors for the $j$-th dimension. We can note that the original bilinear fusion method requires too many parameters and suffers from the heavy computation cost. Thus, we replace it with the low-rank version~\cite{kim2017hadamard}, given by 
\begin{eqnarray}
{\bf f}_i = {\bf P}^{f}(\sigma({\bf W}^v {\bf \widetilde h}^v_i) \odot \sigma({\bf W}^s {\bf \widetilde h}^s_i)) + {\bf b}^f,
\end{eqnarray}
where the ${\bf f}_i$ is the biliner fusion result at the time step $i$.

Eventually, by the attentive aggregation, cross-gated interaction and low-rank bilinear fusion, we obtain the cross-modal semantic representations for each frame, denoted by ${\bf f} = ({\bf f}_1, {\bf f}_2, \ldots, {\bf f}_n)$.

\subsection{Moment Retrieval Module}
In this section, we present the moment retrieval module to simultaneously score a set of candidate moments with multi-scale windows at each time step, and further adopt a temporal boundaries regression mechanism to adjust the moment boundaries.

By the cross-modal interaction module, we get the the cross-modal semantic representations ${\bf f} = ({\bf f}_1, {\bf f}_2, \ldots, {\bf f}_n)$. To absorb the contextual evidences, we likewise develop another BiGRU networks to learn the final semantic representations ${\bf h}^f = ({\bf h}^f_1, {\bf h}^f_2, \ldots, {\bf h}^f_n)$. We then pre-define a set of candidate moments with multi-scale windows at each time step $i$, denoted by  $C_i =\{ ({\hat s}_{ij}, {\hat e}_{ij}) \}_{j=1}^k$, where $({\hat s}_{ij}, {\hat e}_{ij})  = (i - w_j/2, i + w_j/2)$ are the start and end boundaries of the $j$-th candidate moment at time $i$, $w_j$ is the width of $j$-th moment and $k$ is the number of moments. Note that we set the fixed window width $w_j$ for $j$-th candidate moment at every time step. Thus, we can simultaneously produce the confidence scores for these moments at time $i$ by a fully connected layer with sigmoid nonlinearity, given by
\begin{eqnarray}
{\bf cs}_{i} = \sigma({\bf W}^c {\bf h}^f_i + {\bf b}^c)
\end{eqnarray}
where the ${\bf cs}_{i} \in \mathbb{R}^{k}$ represents the confidence scores of $k$ moments at time $i$ and $cs_{ij}$ corresponds to the $j$-th moment.
Likewise, we produce the predicted offsets for these moments by
\begin{eqnarray}
{\hat \delta}_{i} = {\bf W}^o {\bf h}^f_i + {\bf b}^o
\end{eqnarray}
where the ${\hat \delta}_{i} \in \mathbb{R}^{2k}$ represents the predicted offsets of $k$ moments at time $i$ and ${\hat \delta}_{ij} = ({\hat \delta}_s, {\hat \delta}_e) $ corresponds to the $j$-th moment.

\textbf{Alignment loss.}  We first adopt an alignment loss to make the moment aligned to the target moment have high confidence scores and the misaligned moment have low confidence scores. Formally, we first compute the IoU (i.e. Intersection over Union) score $IoU_{ij}$ of each candidate moment $C_{ij}= ({\hat s}_{ij}, {\hat e}_{ij}) $ with the target moment $(s,e)$. If the IoU score of a candidate is less than a clearing threshold $\lambda$, we reset it to 0. Next, we calculate the alignment loss by 
\begin{equation}
\begin{split}
&{\mathcal L}_{ij}  = (1 - IoU_{ij}) \cdot log(1-cs_{ij}) + IoU_{ij} \cdot log(cs_{ij}) \\
&{\mathcal L}_{align} =-\frac{1}{nk}\sum_{i=1}^{n}\sum_{j=1}^{k} {\mathcal L}_{ij} 
\end{split}
\end{equation}
where we consider all candidate moments during alignment training, and apply the concrete IoU score rather than set 0 or 1 according to a threshold value for every candidate. This setting is helpful for distinguishing high-score candidates.

\textbf{Regression loss.} As these multi-scale temporal windows have fixed widths, our candidate moments are restricted to discrete boundaries. To go beyond this limitation, we apply a boundary regression mechanism to adjust the temporal boundaries of high-score moments. Concretely, we fine-tune the localization offsets of high-score moments by a regression loss.
First, we define a set $C_h$ of high-score moments which $IoU$ scores are larger than a high-score threshold $\gamma$, then compute the start and end offset values for those high-score moments as follows:
\begin{eqnarray}
\delta_s = s - {\hat s}, \  \delta_e = e - {\hat e}, 
\end{eqnarray}
where $(s,e)$ are the boundaries of the target moment, and $({\hat s},{\hat e})$ are the boundaries of a high-score moment in $C_h$.
Thus, the $(\delta_s, \delta_e)$ denote its ground truth offsets and the predicted offsets $({\hat \delta}_s, {\hat \delta}_e)$ are given by preceding fully connected layer.
Next, we design the regression loss as follows:
\begin{eqnarray}
{\mathcal L}_{reg} =\frac{1}{N}\sum_{C_h} ( R({\delta}_s - {\hat \delta}_s) + R({\delta}_e - {\hat \delta}_e))
\end{eqnarray}
where $C_h$ is the set of high-score moments, $N$ is the size of $C_h$ and $R$ represents the smooth L1 function.

With the alignment loss and regression loss, we eventually propose a multi-task loss to train the cross-modal interaction networks in an end-to-end manner, denoted by
\begin{eqnarray}
{\mathcal L} = {\mathcal L}_{align} + \alpha {\mathcal L}_{reg},
\end{eqnarray}
where $\alpha$ is a hyper-parameter to control the balance of two losses.

During inference, we simply choose the candidate moment with the highest confidence score. If we need to select multiple moments (i.e. Top K), we first rank all candidates according to their confidence scores and adopt a non-maximum suppression (NMS) to select moments in order.

\section{Experiments}

\subsection{Datasets}
We first introduce two public datasets for query-based moment retrieval.

\textbf{ActivityCaption}~\cite{krishna2017dense}: The ActivityCaption dataset is originally developed for the task of dense video caption, which contains 20,000 untrimmed videos and each video includes multiple natural language descriptions with temporal annotations. The video contents of this dataset are diverse and open. For query-based moment retrieval, each description is regarded as a query and corresponds to a target moment. Since the caption annotations of test data of ActivityCaption are not publically available, we take the val\_1 as the validation set and val\_2 as test data. The details of the ActivityCaption dataset are summarized in Table~\ref{table:dataset}.

\textbf{TACoS}~\cite{regneri2013grounding}: The TACoS dataset is developed onMPII Compositive~\cite{rohrbach2012script} and only contains 127 videos. But each video of TACoS has a large amount of temporal textual annotations. The contents of TACoS are limited to cooking scenes, thus lack the diversity. Moreover, the videos of TACoS are longer but the target moments are shorter than ActivityCaption, which make the query-based moment retrieval harder. The details of this dataset are also summarized in Table~\ref{table:dataset}.

\begin{table}[t]
\centering
\caption{Summaries of ActivityCaption and TACoS Datasets, including number of samples, average video duration, average target moment duration and average query length.}\label{table:dataset}
\begin{tabular}{c|ccccc}
\hline
\hline
 \multicolumn{5}{c}{ActivityCaption}\\
\hline
\hline
     &Number&   Video Time&   Target Time &  Query Len\\
\hline
Train&37,421&117.30&35.45&13.48\\
\hline
Valid&17,505&118.23&37.73&13.58\\
\hline
Test&17,031&118.21&40.25&12.02\\
\hline
All&71,957&117.74&37.14&13.16\\
\hline
\hline
 \multicolumn{5}{c}{TACoS}\\
\hline
\hline
     &Number&    Video Time&   Target Time &  Query Len\\
\hline
Train&10,146&224.16&5.70&8.69\\
\hline
Valid&4,589&387.46&6.23&9.12\\
\hline
Test&4,083&367.70&6.96&9.00\\
\hline
All&18,818&296.21&6.10&8.86\\
\hline
\end{tabular}
\end{table}

\begin{table}[t]
\centering
\caption{Performance Evaluation Results on the ActivityCaption Dataset ($n \in \{1,5\}$ and  $m \in \{0.3,0.5,0.7\}$). }\label{table:activityres}
\scalebox{0.9}{
\begin{tabular}{c|cccccc}
\hline
\hline
 \multirow{2}{*}{Method}&R@1 &R@1 &R@1 &R@5 &R@5 &R@5\\
 &  IoU=0.3&   IoU=0.5 &  IoU=0.7 & IoU=0.3&   IoU=0.5 &  IoU=0.7 \\
\hline
MCN&39.35&21.36&6.43&68.12&53.23&29.70\\
VSA-RNN&39.28&23.43&9.01&70.84&55.52&32.12\\
VSA-STV&41.71&24.01&8.92&71.05&56.62&34.52\\
CTRL&47.43&29.01&10.34&75.32&59.17&37.54\\
ACRN&49.70&31.67&11.25&76.50&60.34&38.57\\
QSPN&52.13&33.26&13.43&77.72&62.39&40.78\\
\hline
CMIN&{\bf 63.61}&{\bf43.40}&{\bf23.88}&{\bf80.54}&{\bf67.95}&{\bf50.73}\\
\hline
\end{tabular}
}
\end{table}

\subsection{Implementation Details}
In this section, we introduce some implementation details of our CMIN method, including the data preprocessing and model setting.

\textbf{Data Preprocessing.} We first resize every frame of videos to 112 $\times$ 112 and extract the visual features by a pre-trained 3D-ConvNet~\cite{tran2015learning}. Specifically, we define continuous 16 frames as a unit and each unit overlaps 8 frames with adjacent units. We then input the units to the pre-trained 3D-ConvNet and obtain 4,096 dimension features for each unit. We next reduce the dimensionality of features from 4,096 to 500 using PCA, which is helpful for decreasing model parameters. These 500-d features are used as the frame features of our CMIN. Since some videos are overlong, we uniformly downsample their feature sequences to 200.

For natural language queries, we first extract the syntactic dependency graph using the library of NLTK~\cite{bird2004nltk} and employ the pre-trained Glove word2vec~\cite{pennington2014glove} to extract the embedding features for each word token. The dimension of word features is 300.

\textbf{Model Setting.} In our CMIN, we sample $k$ candidate moments with multi-scale windows at each time step.
Concretely, we set $7$ window widths of $[16,32,64,96,128,160,196]$ for the ActivityCaption dataset and $4$ window widths of $[8,16,32,64]$  for TACoS. Thus, we have 1,400 samples for each video on ActivityCaption and 800 samples on TACoS. Note that we cut off candidate examples that are beyond the boundaries of videos. 
We then set the clearing threshold $\lambda$ to 0.3, the high-score threshold $\gamma$ to 0.7, and the balance hyper-parameter $\alpha$ to 0.001. Moreover, the dimension of the hidden state of BiGRU networks is set to 512 (256 for one direction). The dimensions of the linear matrice in the multi-head self-attention and bilinear fusion are also set to 512.
During training, we adopt an adam optimizer~\cite{duchi2011adaptive} to minimize the multi-task loss and the learning rate is set to 0.001. We employ a mini-batch method and the batch size is 128. 

\subsection{Evaluation Criteria}
To measure the retrieval performance of our CMIN and baselines, we adopt the ``R@n, IoU=m'' as evaluation criteria, which are proposed in~\cite{gao2017tall}. Concretely, we first calculate the IoU (i.e. Intersection over Union) between the selected moment and ground-truth moment, and the ``R@n, IoU=m'' means the percentage of at least one of top-n selected moments having IoU larger than m. The metric is on the query level, so the overall performance is the average among all the queries, denoted by $R(n,m) = \frac{1}{N_q} \sum_{i=1}^{N_q}r(n,m,q_i)$, where the $r(n,m,q_i)$ represents whether one of the top-n selected moments of the query $q_i$ has $IoU > m$, and $N_q$ is the total number of testing queries.

\begin{table}[t]
\centering
\caption{Performance Evaluation Results on the TACoS Dataset ($n \in \{1,5\}$ and  $m \in \{0.1, 0.3,0.5\}$).}\label{table:tacosres}
\scalebox{0.9}{
\begin{tabular}{c|cccccc}
\hline
\hline
 \multirow{2}{*}{Method}&R@1 &R@1 &R@1 &R@5 &R@5 &R@5\\
 &  IoU=0.1&   IoU=0.3 &  IoU=0.5 & IoU=0.1&   IoU=0.3 &  IoU=0.5 \\
\hline
MCN&3.11&1.64&1.25&3.11&2.03&1.25\\
VSA-RNN&8.84&10.77&4.78&19.05&13.90&9.10\\
VSA-STV&15.01&10.77&7.56&32.82&23.92&15.50\\
CTRL&24.32&18.32&13.30&48.73&36.69&25.42\\
ACRN&24.22&19.52&14.62&47.42&34.97&24.88\\
QSPN&25.31&20.15&15.23&53.21&36.72&25.30\\
\hline
CMIN&{\bf 32.48}&{\bf24.64}&{\bf18.05}&{\bf62.13}&{\bf38.46}&{\bf27.02}\\
\hline
\end{tabular}
}
\end{table}

\subsection{Performance Comparisons}

We compare our proposed CMIN method with some existing state-of-the-art methods to verify the effectiveness.
\begin{itemize}
\item \textbf{MCN}~\cite{hendricks2017localizing}: The MCN method adopts a moment context network to integrate local and global moment features for query-based moment retrieval.
\item \textbf{VSA-RNN and VSA-STV}~\cite{gao2017tall}: The two methods are the extensions of the DVSA model~\cite{karpathy2015deep}. They both simply transforms the visual features of candidate moments and query features into a common space, and then estimate the correlation scores to select the most relevant one. The VSA-RNN applies a LSTM network to encode queries and VSA-STV adopts the off-the-shelf skip-thought~\cite{kiros2015skip} feature extractor.
\item \textbf{CTRL}~\cite{gao2017tall}: The CTRL method proposes a cross-modal temporal regression localizer to estimate the alignment scores of candidate moments and textual queries by leveraging contextual contents of these moments, and then adjust the start and end boundaries of high-score moments. 
\item \textbf{AMRN}~\cite{liu2018attentive}: The AMRN method emphasizes the visual moment features by attentive contextual contents and develops a cross-modal feature representation.
\item \textbf{QSPN}~\cite{xu2019multilevel}: The QSPN method introduces a multi-level model to integrate visual and textual features earlier with an attention mechanism, learn spatio-temporal visual representations and further re-generate queries as the auxiliary task.
\end{itemize}
The former three approaches only focus on the visual features within each moment and ignore the contextual information. And the latter three approaches incorporate the contextual evidence to improve the retrieval performance, where the AMRN utilizes the attention mechanism to filter irrelevant context and QSPN further develops an early interaction strategy for cross-modal features.

Table~\ref{table:activityres} and Table~\ref{table:tacosres} show the overall performance evaluation results of our CMIN and all baselines on ActivityCaption and TACoS datasets, respectively. We choose the evaluation criteria ``R@n, IoU=m'' with $n \in \{1,5\}$, $m \in \{0.3,0.5,0.7\}$ for ActivityCaption and $n \in \{1,5\}$ , $m \in \{0.1,0.3,0.5\}$ for TACoS. Note that we report the baseline performance based on either their original paper or our implementation by selecting the higher one.
The experimental results reveal a number of interesting points:
\begin{itemize}
\item While modeling moment features, the MCN applies a mean-pooling operation to aggregate all features of video sequences as context of the current moment, which may introduce noises into the moment representations and degrade the retrieval accuracy. Thus, the MCN achieves the worst performance on all criteria.
\item The context based methods CTRL, ACRN, QSPN and CMIN outperform the simple SVA-STV and VSA-RNN, which suggest the context modeling is crucial for high-quality moment retrieval. And the performance of the VSA-STA is slightly better than VSA-RNN, demonstrating the skip-thought feature extractor is helpful for query understanding.
\item The QSPN adopts an attention mechanism to fuse the visual and textual features, which develop the early interactions between moments and queries. The fact that the QSPN achieves better performance than CTRL and ACRN verifies the cross-modal interaction is critical for query-based moment retrieval.
\item On all the criteria of two datasets, the CMIN not only outperforms all previous state-of-the-art baselines, but also achieves tremendous improvements, especially on ActivityCaption. These results verify the effectiveness of our syntactic GCN, multi-head self-attention and multi-stage cross-modal interaction.
\end{itemize}

Moreover, we can find that the overall experimental results on TACoS are lower than ActivityCaption, and meanwhile, the CMIN can only achieve a smaller improvement on TACoS. That is because the videos are longer and the target moments are shorter on TACoS as shown in Table~\ref{table:dataset}, which increase the number of potential candidate moments and make this task harder. Moreover, the invariant cooking scenes and shorter query description may also improve the retrieval difficulty. Since invariant scenes require better discrimination ability and the short query is hard to describe a moment clearly.

\subsection{Ablation Study}
To prove the contribution of each component of our CMIN method, we next conduct some ablation studies on the syntactic GCN, multi-head self-attention and multi-stage cross-modal interaction. Concretely, we discard one component at a time to generate an ablation model as follows.
\begin{itemize}
\item \textbf{CMIN(w/o. GCN)}: We first remove the syntactic GCN layer from the query representation learning and take the query representations from BiGRU networks as the input of multi-stage cross-modal interaction module.
\item \textbf{CMIN(w/o. SA)}: We then discard the multi-head self-attention from the video representation learning to validate the importance of long-range semantic dependency modeling.
\item \textbf{CMIN(w/o. CG)}: We next remove the cross gate in the multi-stage cross-modal interaction module, directly applying the bilinear fusion for frame representations and aggregation query representations.
\item \textbf{CMIN(w/o. BF)}: We finally replace the low-rank bilinear fusion method with a simple concatenation for query and video features.
\end{itemize}
The ablation results on ActivityCaption and TACoS datasets are shown in Table~\ref{table:activityablation} and Table~\ref{table:tacosablation}, respectively. By analyzing the ablation results, we can find several interesting points:
\begin{itemize}
\item The CMIN(full) outperforms all ablation models on both ActivityCaption and TACoS datasets, which demonstrates the syntactic GCN, multi-head self-attention, cross gate and low-rank bilinear fusion are all helpful for query-based moment retrieval.
\item The CMIN(w/o. GCN) achieves the worst performance on AcitivityCaption, and also have poor results on TACoS, indicating the utilization of syntactic structure is critical for query semantic understanding and subsequent modeling.
\item All ablation models still yield better results than all baselines. This fact demonstrates that our comprehensive retrieval framework is suitable for this task and the excellent performance does not only depend on a key component.
\end{itemize}

\begin{table}[t]
\centering
\caption{Performance Evaluation Results of Ablation Model on the ActivityCaption dataset.}\label{table:activityablation}
\scalebox{0.9}{
\begin{tabular}{c|cccccc}
\hline
\hline
 \multirow{2}{*}{Method}&R@1 &R@1 &R@1 &R@5 &R@5 &R@5\\
 &  IoU=0.3&   IoU=0.5 &  IoU=0.7 & IoU=0.3&   IoU=0.5 &  IoU=0.7 \\
\hline
w/o. GCN&60.12&40.84&21.79&78.23&65.67&45.43\\
w/o. SA&61.22&41.56&22.36&79.43&66.91&48.12\\
w/o. CG&60.57&41.21&22.01&78.62&65.99&46.89\\
w/o. BF&61.32&41.89&22.12&79.27&66.21&47.92\\
\hline
full&{\bf 63.61}&{\bf43.40}&{\bf23.88}&{\bf80.54}&{\bf67.95}&{\bf50.73}\\
\hline
\end{tabular}
}
\end{table}

\begin{table}[t]
\centering
\caption{Performance Evaluation Results of Ablation Model on the TACoS dataset.}\label{table:tacosablation}
\scalebox{0.9}{
\begin{tabular}{c|cccccc}
\hline
\hline
 \multirow{2}{*}{Method}&R@1 &R@1 &R@1 &R@5 &R@5 &R@5\\
 &  IoU=0.1&   IoU=0.3 &  IoU=0.5 & IoU=0.1&   IoU=0.3 &  IoU=0.5 \\
\hline
w/o. GCN&30.54&23.22&10.03&57.69&37.12&26.16\\
w/o. SA&30.21&23.02&16.87&55.54&36.6&25.37\\
w/o. CG&31.96&23.59&17.47&61.87&38.11&26.79\\
w/o. BF&32.01&{\bf 24.79}&17.61&61.59&38.23&26.75\\
\hline
full&{\bf 32.48}&24.64&{\bf18.05}&{\bf62.13}&{\bf38.46}&{\bf27.02}\\
\hline
\end{tabular}
}
\end{table}

\begin{figure}[t]
\centering
\includegraphics[width=0.23\textwidth]{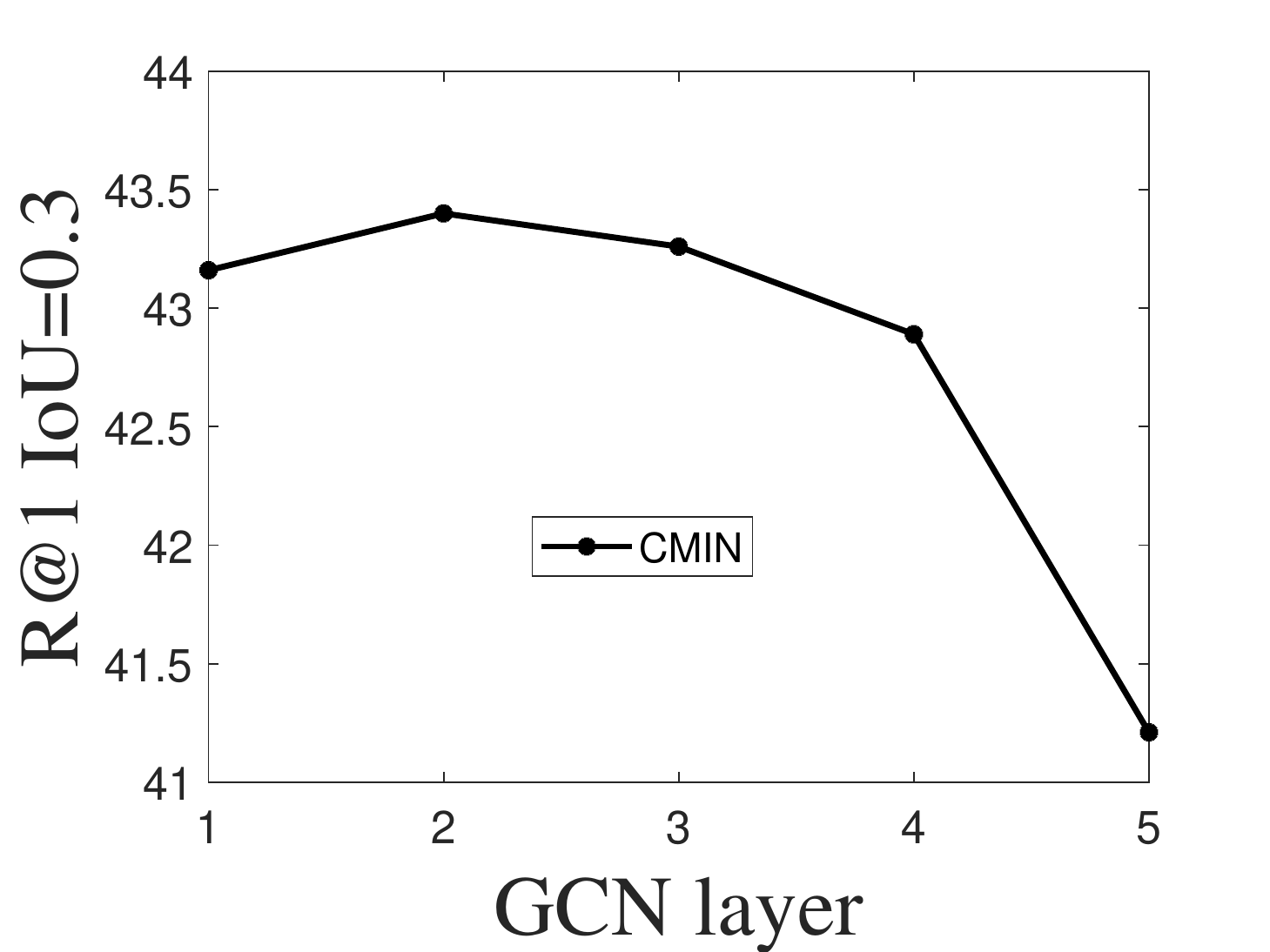}
\includegraphics[width=0.23\textwidth]{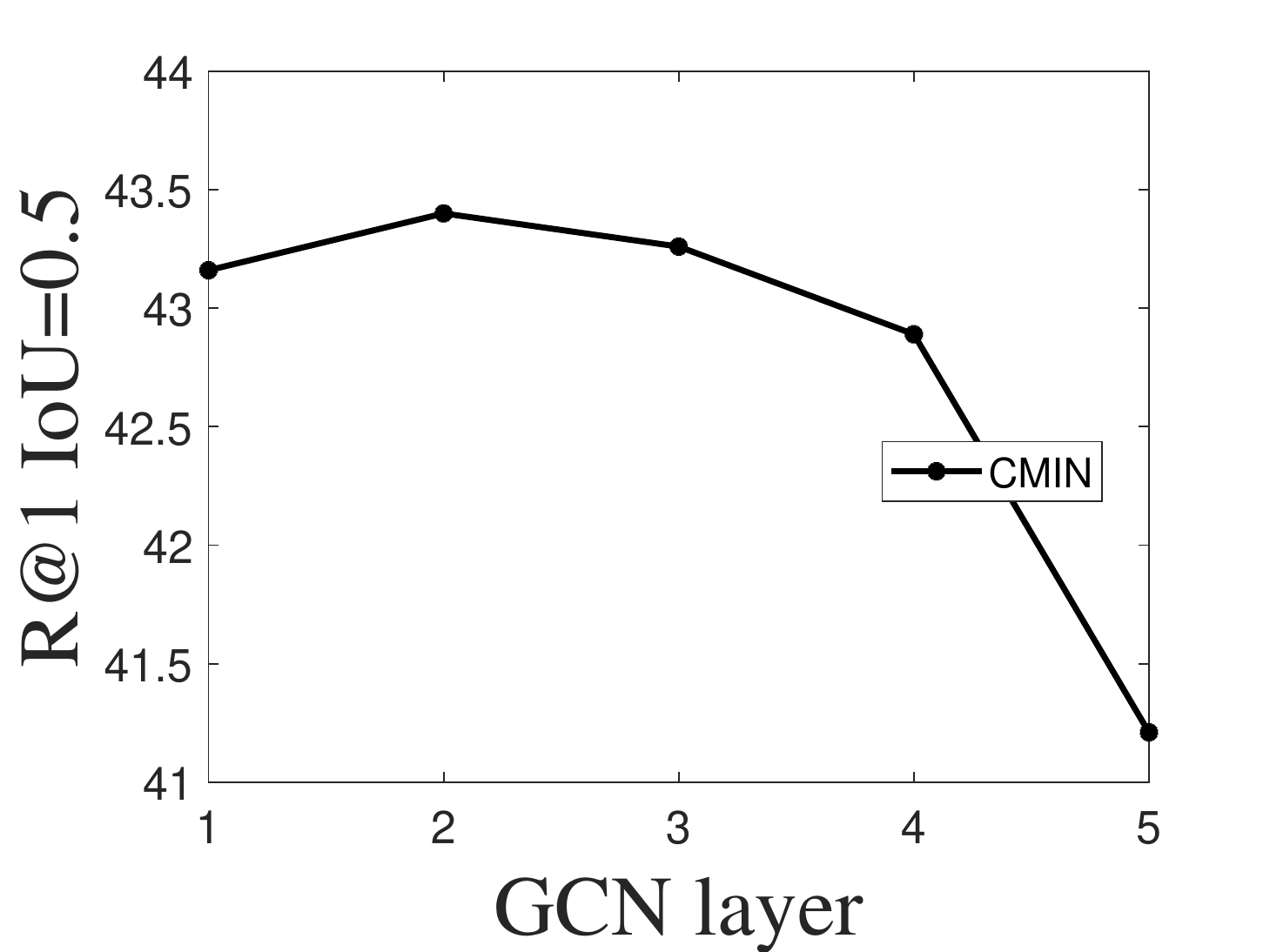}
\caption{Effect of the Number of Stacked Syntactic GCN layers on the ActivityCaption Dataset.}\label{fig:activitylayer}
\end{figure}

\begin{figure}[t]
\centering
\includegraphics[width=0.23\textwidth]{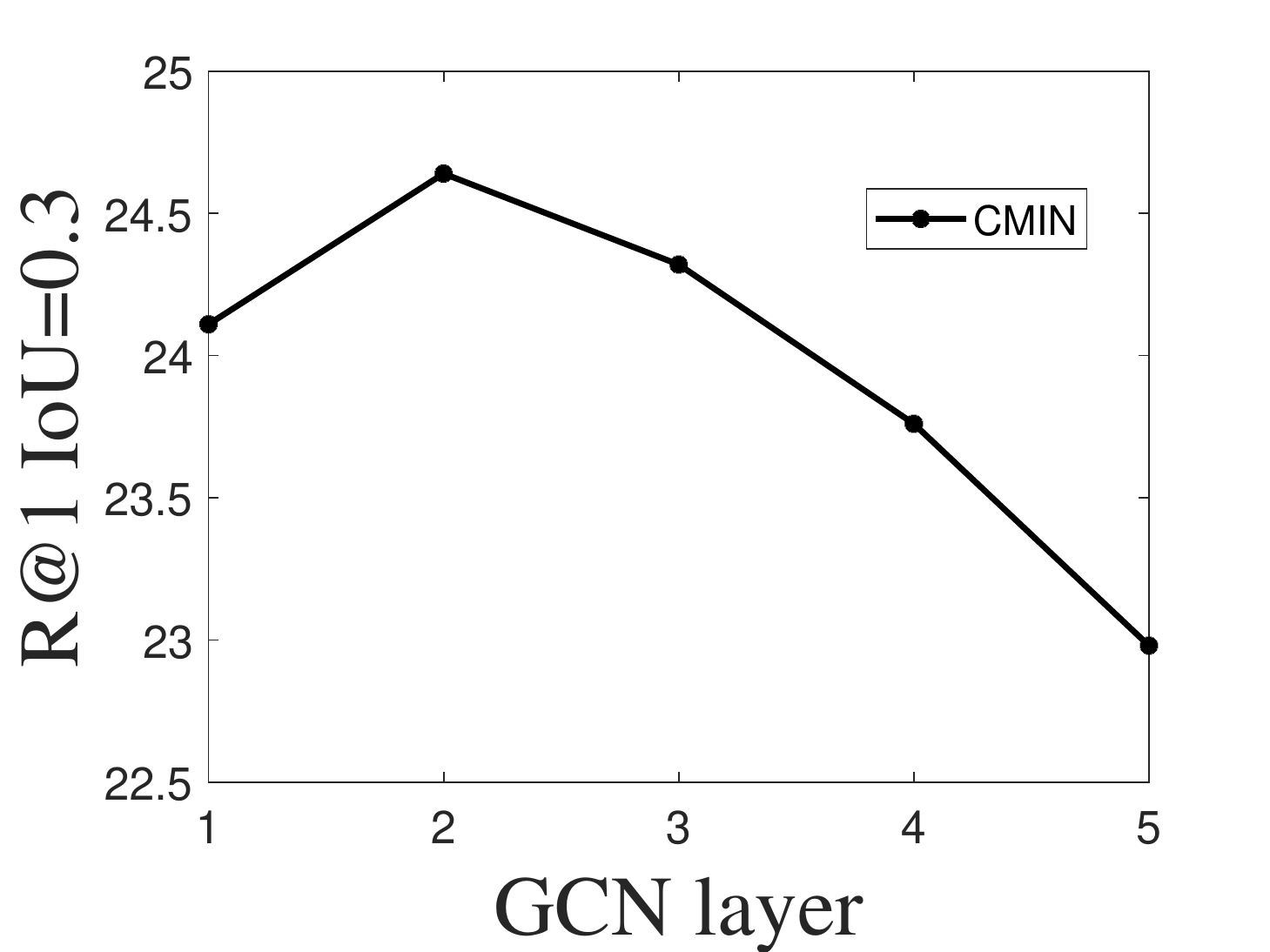}
\includegraphics[width=0.23\textwidth]{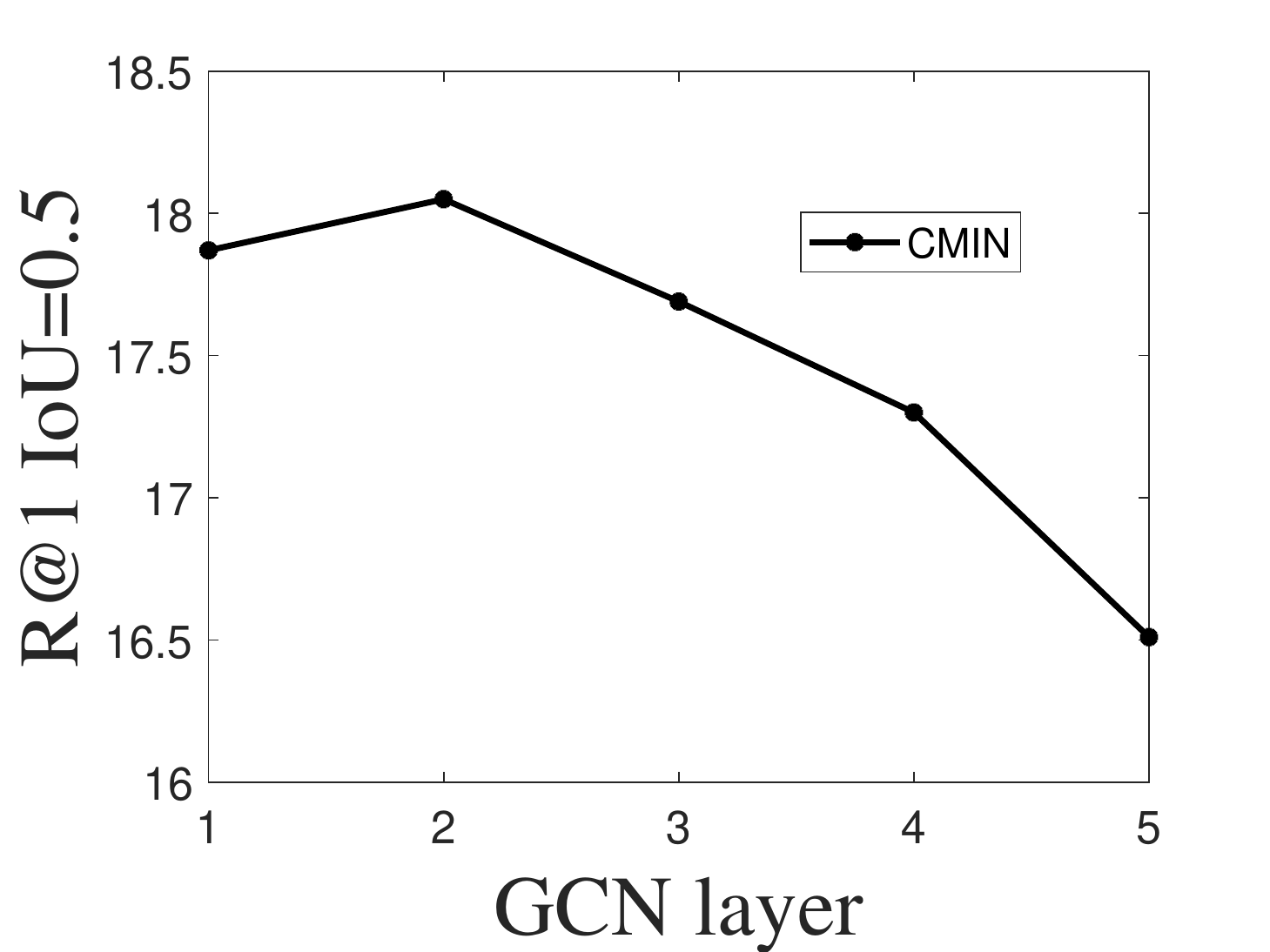}
\caption{Effect of the Number of Stacked Syntactic GCN layers on the TACoS Dataset.}\label{fig:tacoslayer}
\end{figure}

Moreover, for the syntactic GCN module, the number of stacked layers is a crucial hyper-parameter. Therefore, We further explore the effect of this hyper-parameter by varying the number of layers from 1 to 5. 
Figure~\ref{fig:activitylayer} and Figure~\ref{fig:tacoslayer} shows the impact of layer number on ActivityCaption and TACoS datasets. Here we select ``R@1,IoU=0.3'' and  ``R@1,IoU=0.5'' as evaluation criteria.
From the tables, we note that the CMIN achieves the best performance while the number of layers is set to 2, and stacking too many or too few layers will both affect the performance of query-based moment retrieval. Because only one syntactic GCN layer cannot sufficiently leverage the syntactic dependencies of natural language queries and too many syntactic GCN layers will result in over-smoothing, that is, each word representation converges to the same value.

\subsection{Qualitative Analysis}
\begin{figure}[t]
\centering
\includegraphics[width=0.48\textwidth]{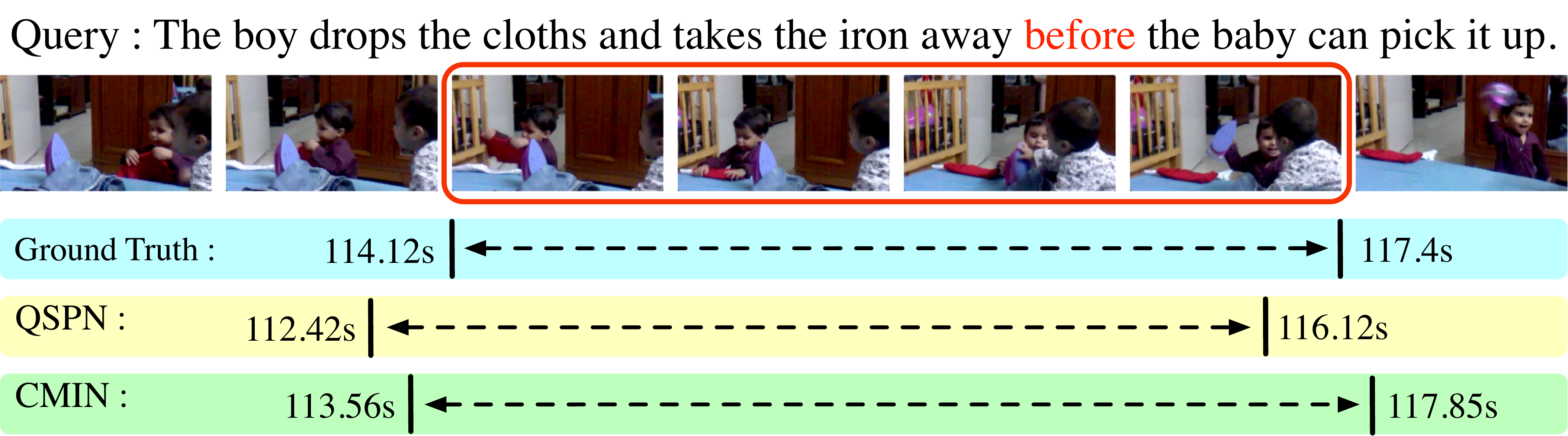}
\includegraphics[width=0.48\textwidth]{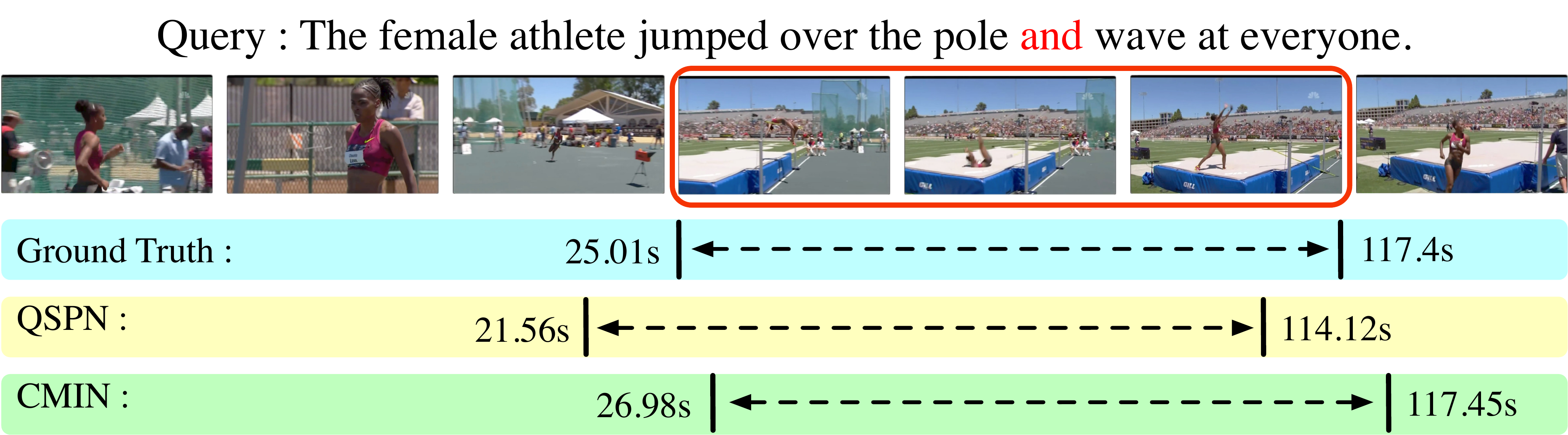}
\caption{Examples on the ActivityCaption dataset.}\label{fig:activityexample}
\end{figure}

\begin{figure}[t]
\includegraphics[width=0.48\textwidth]{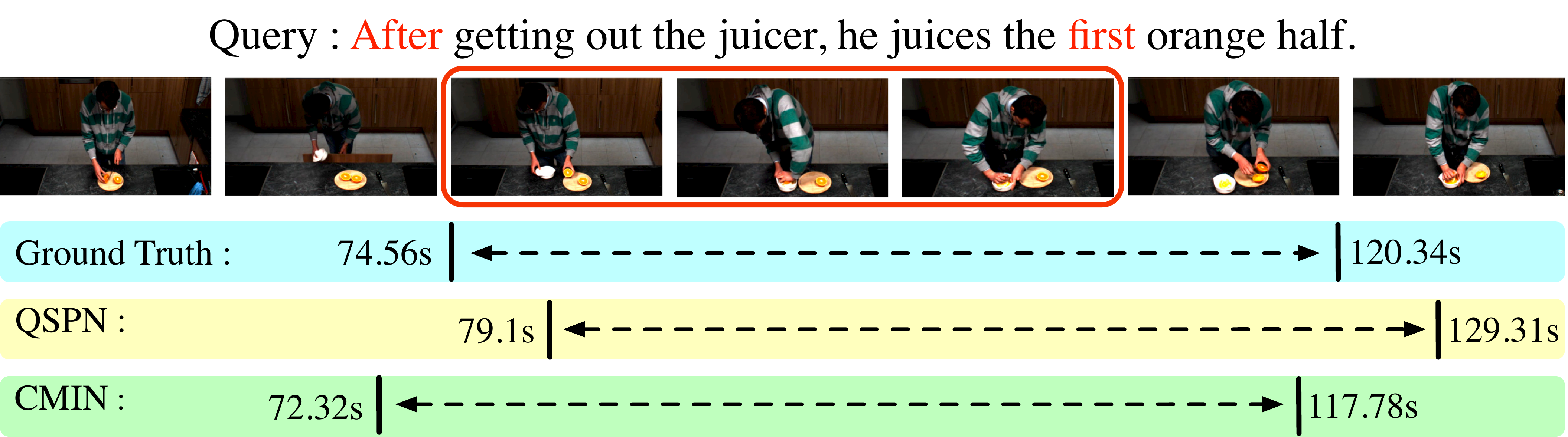}
\includegraphics[width=0.48\textwidth]{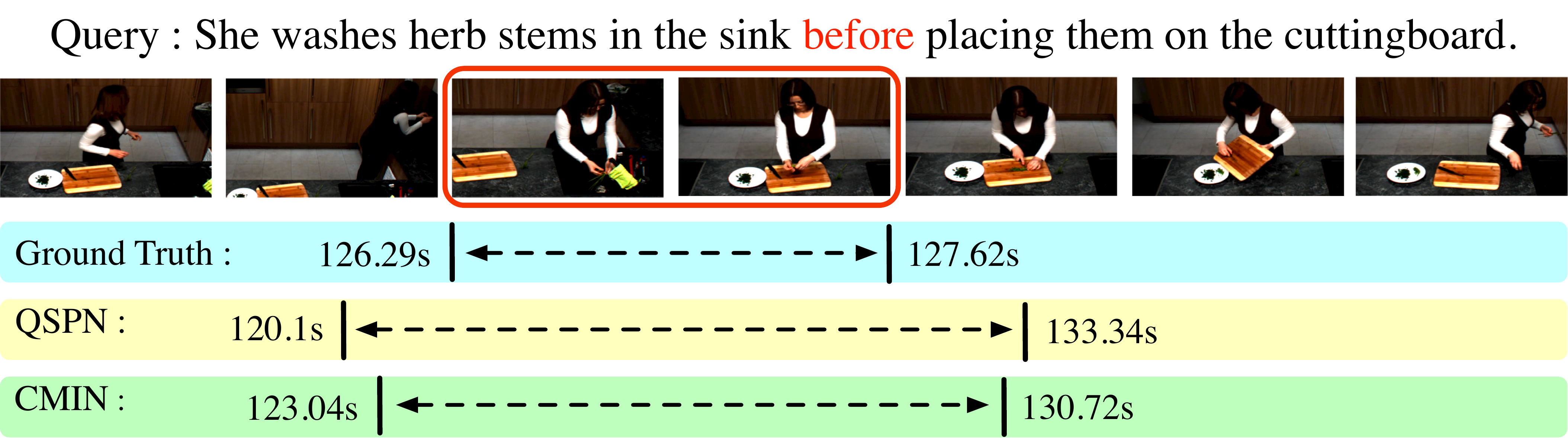}
\caption{Examples on the TACoS dataset.}\label{fig:tacosexample}
\end{figure}

To qualitatively validate the effectiveness of the CMIN method, we display several typical examples of query-based moment retrieval.
Figure~\ref{fig:activityexample} and Figure~\ref{fig:tacosexample} show the retrieval results of the CMIN method and the best baseline QSPN on ActivityCaption and TACoS datasets, respectively.
We can find that natural language queries are very diverse and often contain successive temporal actions. By intuitive comparison, the CMIN can retrieve more accurate boundaries of target moments than QSPN. Moreover, the retrieval precision on TACoS is lower than ActivityCaption, which is consistent with previous qualitative evaluations.

\begin{figure}[t]
\includegraphics[width=0.48\textwidth]{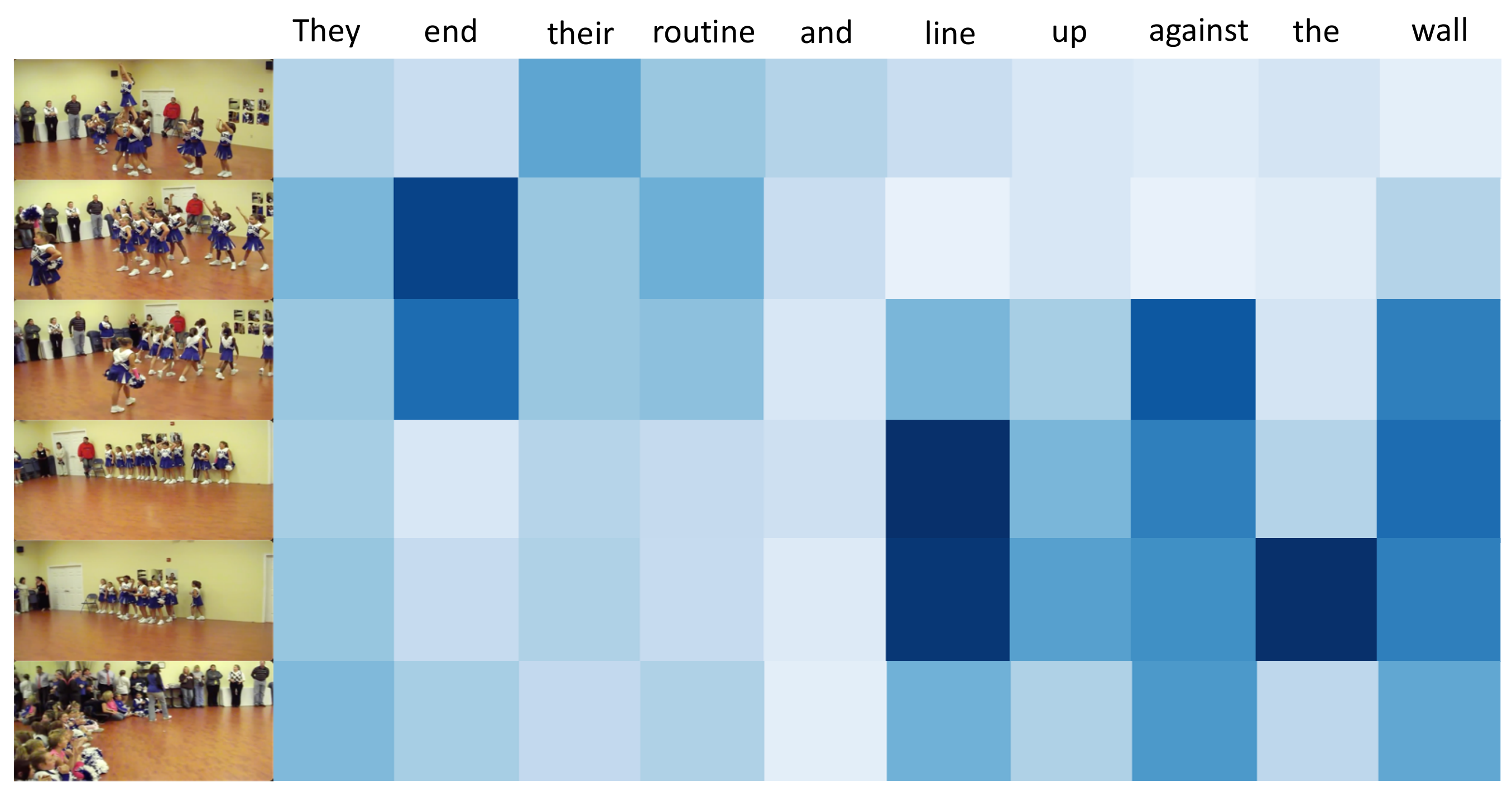}
\caption{The Video-to-Query Attention Results in the Multi-Stage Cross-Modal Interaction Module}\label{fig:attention}
\end{figure}

Furthermore, as the fundamental component of our multi-stage cross-modal interaction module, the video-to-query attentive aggregation builds a bridge between video and query information. Thus, we demonstrate how the attentive aggregation mechanism works to further understand the interaction process.
As shown in Figure~\ref{fig:attention}, the video-to-query attention results are visualized using a thermodynamic diagram, where the darker color means the higher correlation of the pair of frame and word representations.
We note that each frame can attend the semantically related words and ignore these irrelevant words.
For example, the word ``line'' has the highest attention score over the query for the fourth frame. This suggests the attentive aggregation strategy effectively establishes the relationship between visual and textual information, and is helpful for high-quality moment retrieval.

\section{Conclusion}
In this paper, we propose a novel cross-modal interaction network for query-based moment retrieval, which considers three critical factors of this task, including the syntactic structure of natural language queries, long-range semantic dependencies in video context and the fine-grained cross-modal interaction.
Specifically, we advise a syntactic GCN to leverage the syntactic structure of queries for fine-grained representation learning, then propose a multi-head self-attention to capture long-range semantic dependencies from video context, and employ a multi-stage cross-modal interaction to explore the potential relations of video and query contents.
The extensive experiments on ActivityCaption and TACoS datasets demonstrate the effectiveness of our proposed method.

%
% The acknowledgments section is defined using the "acks" environment (and NOT an unnumbered section). This ensures
% the proper identification of the section in the article metadata, and the consistent spelling of the heading.
\begin{acks}
This work was supported by the National Natural Science Foundation of China under Grant No.61602405, No.U1611461, No.61751209 and No.61836002, Sponsored by Joint Research Program of ZJU and Hikvision Research Institute.
\end{acks}

%
% The next two lines define the bibliography style to be used, and the bibliography file.
\bibliographystyle{ACM-Reference-Format}
\balance
\bibliography{sigir19-cmin}

% 
% If your work has an appendix, this is the place to put it.
% \appendix



% Lorem ipsum dolor sit amet, consectetur adipiscing elit. Morbi malesuada, quam in pulvinar varius, metus nunc fermentum urna, id sollicitudin purus odio sit amet enim. Aliquam ullamcorper eu ipsum vel mollis. Curabitur quis dictum nisl. Phasellus vel semper risus, et lacinia dolor. Integer ultricies commodo sem nec semper. 


% Etiam commodo feugiat nisl pulvinar pellentesque. Etiam auctor sodales ligula, non varius nibh pulvinar semper. Suspendisse nec lectus non ipsum convallis congue hendrerit vitae sapien. Donec at laoreet eros. Vivamus non purus placerat, scelerisque diam eu, cursus ante. Etiam aliquam tortor auctor efficitur mattis. 


% Nam id fermentum dui. Suspendisse sagittis tortor a nulla mollis, in pulvinar ex pretium. Sed interdum orci quis metus euismod, et sagittis enim maximus. Vestibulum gravida massa ut felis suscipit congue. Quisque mattis elit a risus ultrices commodo venenatis eget dui. Etiam sagittis eleifend elementum. 

% Nam interdum magna at lectus dignissim, ac dignissim lorem rhoncus. Maecenas eu arcu ac neque placerat aliquam. Nunc pulvinar massa et mattis lacinia.

\end{document}